\documentclass[aps,pre,twocolumn,groupedaddress,showpacs]{revtex4}
\usepackage{graphicx}
\usepackage{CJK}
\newcommand{\tri}{\triangle}

\newcommand{\br}{\mathbf{r}}

\newcommand{\sep}{ \ \ \ , \ \ \ }

\newcommand{\beq}{\begin{equation}}
\newcommand{\eeq}{\end{equation}}
\newcommand{\beqn}{\begin{eqnarray}}
\newcommand{\eeqn}{\end{eqnarray}}
\newcommand{\pp}{\partial}

\newcommand{\eq}{Eq.\ }

\newcommand{\vnabla}{\vec{\nabla}}

\newcommand{\fig}{{Fig.\ }}
\newcommand{\sect}{{Section }}
\newcommand{\tab}{{Table }}

\newcommand{\ch}{Ch.\ }

\begin{document}
\begin{CJK*}{CNS1}{}


\title{Elongation dynamics of amyloid fibrils: a rugged energy landscape picture}
\author{Chiu Fan \surname{Lee}
(\CJKchar[CNS1]{"4A}{"57} \CJKchar[CNS1]{"62}{"3E} \CJKchar[CNS1]{"47}{"7D})
}
\email{C.Lee1@physics.ox.ac.uk}
\affiliation{Physics Department, Clarendon Laboratory,
Oxford University, Oxford OX1 3PU, UK}
\author{James Loken}
\affiliation{Physics Department, Denys Wilkinson Building,
Oxford University, Oxford OX1 3RH, UK}
\author{Letitia Jean}
\affiliation{Sir William Dunn School of Pathology,
Oxford University, Oxford OX1 3RE, UK}
\author{David J. Vaux}
\affiliation{Sir William Dunn School of Pathology,
Oxford University, Oxford OX1 3RE, UK}
\date{\today}

\begin{abstract}
Protein amyloid fibrils are a form of linear protein aggregates that are implicated in many neurodegenerative diseases.
Here,
we study the dynamics of amyloid fibril elongation by performing Langevin dynamic simulations on a coarse-grained model of peptides. Our simulation results suggest that the elongation process is dominated by a series of local minimum due to frustration in monomer-fibril interactions. This rugged energy landscape picture indicates that the amount of recycling of monomers at the fibrils' ends before being fibrillized is substantially  reduced in comparison to the conventional two-step elongation model. This picture, along with other predictions discussed, can be tested with current experimental techniques.
\end{abstract}


\maketitle
\end{CJK*}

\section{Introduction}
Amyloids are insoluble fibrous protein aggregations stabilized by a network of 
hydrogen bonds and hydrophobic interactions \cite{Sunde_JMB97, Dobson_Nature03, 
Radford_TrendsBiochemSci00,Sawaya_Nature07}.
They are intimately related to many neurodegenerative diseases such as
Alzheimer's Disease, Parkinson's Disease and prion diseases 
\cite{Harper_AnnuRevBiochem97}.
Better characterization of the various properties of amyloid fibrils is 
therefore of high importance for the understanding of the associated 
pathogenesis.
In this paper, we investigate the dynamics of elongation in fibril growth.

\begin{figure}
\caption{
Schematic pictures depicting the model adopted in this work.
(a) Each amino acid is simplistically represented by two beads, the gray beads represent the peptide backbone of an amino acid and the coloured beads the side chains (red for hydrophobic and green for hydrophilic).
We stress that the representation is only meant to be qualitative.
(b) The five-amino-acid peptide employed in this work with alternating hydrophilic-hydrophobic side chains. The alternating pattern has been shown to promote amyloid fibril formation \cite{Yoon_ProtSci04,
Fernandez_NatBiotech04,Tartaglia_ProtSci05,Galzitskaya_PLoSCompBiol06,DuBay_JMB04,Pawar_JMB05}.
(c) A segment of fibril consisting of four peptides in two layers of cross-beta sheets.
(d) A cartoon depicting the four beta strands corresponding to (c). The green (red) face of the panel depicts the hydrophilic (hydrophobic) side.
}
\label{main_pic}
\begin{center}
\includegraphics[scale=.45]{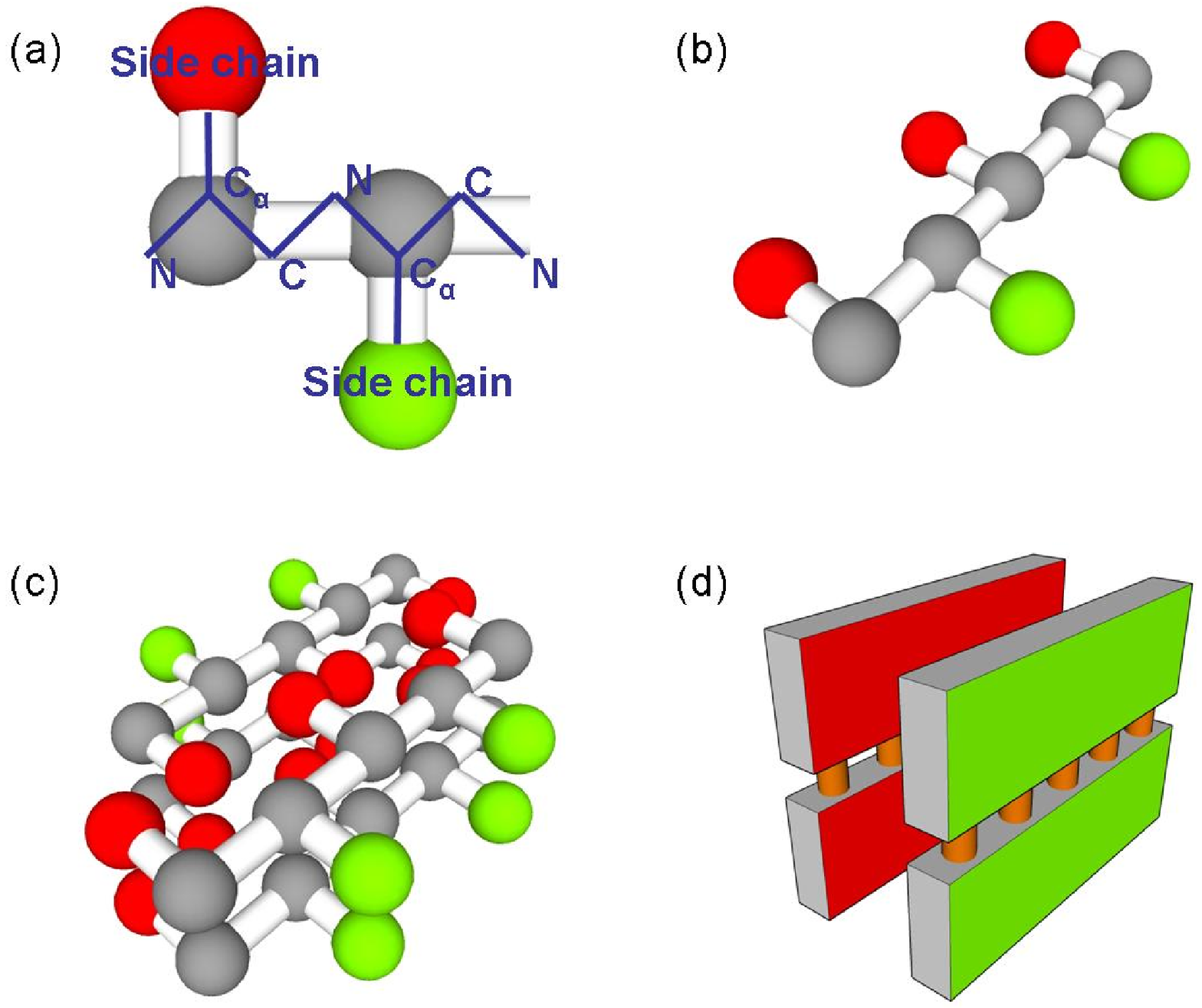}
\end{center}
\end{figure}

From the kinetic theory point of view, the elongation process is traditionally viewed
as a diffusion-limited reaction coupled with high free-energy barrier crossing \cite{Hall_JMB05}.
  A more refined picture has also been proposed in \cite{Kusumoto_PNAS98,Esler_Biochem00,Massi_Proteins00,Scheibel_PNAS04,Collins_PLoSBiol04,Nguyen_PNAS07} where the elongation process is treated as a two-step dock-and-convert process. Namely, the elongation process follows the kinetics scheme:
\beq
\label{scheme}
[m] +[f_k]
\begin{array}{c}
a_+
\\
\rightleftharpoons
\\
a_-
\end{array}
[(m\oplus f_k)]
\begin{array}{c}
b_+
\\
\rightharpoonup
\\
\
\end{array}
[f_{k+1}]
\eeq
where $[m]$ denotes the monomer concentration, $[f_k]$ the fibril  concentration consisting of $k$ monomers, $[(m\oplus f_k)]$ denotes the intermediate state before a monomer can be converted into fibril, and $a_+$, $a_-$ and $b$ are the corresponding rates. Intuitively, the intermediate state $(m\oplus f_k)$ corresponds to the initial state when the monomer  first becomes bound to the fibril's end through hydrophobic interactions or/and a small number of hydrogen bonds. This suggests that at the initial binding, the free energy gain should amount to only a few $k_BT$ \cite{Sneppen_B05}. Taking $a_+$ to be about $10^{10}$M$^{-1}$s$^{-1}$, which is the typical rate for a diffusion-limited bio-molecular reaction \cite{Jackson_B06}, and the binding free energy to be $3k_BT$, say, then $a_-$ can be estimated from the law of mass actions as: $a_- \sim a_+ \exp(-3)\times$M  $\sim 10^8$s$^{-1}$ \footnote{This estimation assumes that the solution is close to being ideal (c.f. \ch 10 in \cite{Landau_B80}).}

Experimentally, elongation rates for amyloid fibrils formed from various peptides have been measured (e.g., \cite{Lomakin_PNAS97, Ban_JMB04, Scheibel_PNAS04, Collins_PLoSBiol04}). In particular, Abeta fibrils, which are implicated in the pathogenesis of Alzheimer's disease \cite{Harper_AnnuRevBiochem97}, have an estimated conversion-limiting elongation rate of
$\sim$0.3$\mu$m/min \cite{Ban_JMB04}. Given that each beta strand is about 0.5nm in width, this elongation rate translates to about one monomer fibrillized per second.
Now, if we assume that $a_- \sim 10^8$s$^{-1}$ as mentioned before, the two-step model depicted in \eq \ref{scheme} predicts that almost $10^8$ monomers would have been interacted with the fibril's end before one of them is converted \footnote{\label{footnote1}This estimation comes from the property that the minimum of two variables drawn from two independent exponential distributions with rates $\lambda_1$ and $\lambda_2$ is again exponential distributed with rates $\lambda_1+\lambda_2$, and that the probability of having the minimum drawn from the first distribution is $\lambda_1/(\lambda_1+\lambda_2)$.}.

In this work, we demonstrate, with the help of coarse-grained molecular dynamics simulations,  that the dynamics in the conversion step is dominated by a series of energy traps manifested by the frustrated monomer-fibril interactions, which would include i) the misalignment of the monomer with respect to the beta strands at the fibril's end; ii) frustrated hydrophobic interactions among the side chains; and iii) the competition to bind between multiple monomers at the same end of a growing fibril. In this picture, there is not a rate limiting step for the elongation process, but rather, each energy trap contributes to the final elongation rate observed. This scenario is akin to the random energy model investigated in glassy systems (e.g., see \cite{Monthus_JPA96}). This rugged energy landscape picture predicts that monomers would spend a substantial amount of time at the fibril's end before conversion. As a result, the amount of recycling of the monomers at the fibrils' ends before one of them becomes fibrillized would be many orders of magnitude small than the value indicated by  the two-step model depicted in \eq \ref{scheme}. This dynamical picture is testable by, e.g., performing fibril elongation experiments with a small portion of monomers radioactively tagged \cite{Collins_PLoSBiol04, Scheibel_PNAS04}.

We will now present the details on the simulation method is presented in \sect \ref{sim}. In \sect \ref{data}, we explain how the simulation results are analyzed. In \sect \ref{discussion}, we discuss how our findings relate to the rugged energy picture introduced and present some implications of our model. We then end with a conclusion.

\section{Simulation details}
\label{sim}
Employing coarse-grained peptide models for the study of amyloids are abundant in the literature (e.g., \cite{Peng_PRE04,Nguyen_PNAS04,Santini_JACS04,Urbanc_PNAS04,Favrin_BiophysJ04,Pellarin_JMB06,Fawzi_JMB07,Li_JCP08}),
here we aim to use the simplest model to study amyloid fibril elongation. Specifically, we keep only two types of inter-peptide interactions: directional interactions provided by hydrogen bonds, and undirectional interactions given by hydrophobic interactions. The directional interactions dictate that fibrils can only grow linearly, and the undirectional interactions are indispensable in keeping the fibrils thermodynamically stable \cite{Lee_2008}. The directionality of the hydrogen bonds suggests that each amino acid has to be represented by at least two set of coordinates, one for its position and one for the direction of the side-chain. We therefore minimally employ two beads to represent each amino-acid (c.f. \fig \ref{main_pic}a). We stress that
we do not attempt to devise a quantitatively correct representation of a peptide, but rather to use a toy model to study the elongation process.

\begin{figure}
\caption{The bonded interactions for the model.
(a) Besides the white bonds between the beads, the orange bonds are employed to fix the angle between the side-chains and the peptide backbone.
Note that the side-chain at the end of the peptide, $B_5$, is bonded to $A_4$ in order to maintain the angle
$B_5-A_5-A_4$. (b) The six bonding interactions in the $D$-bond between beads $A_k$ and $A_h$. If one of the $A$ beads involved is at the end of the peptide, e.g., $A_h=A_5$, then the cross-peptide $A$-$A$ bonds will be between are $A_k-A_5$, $A_{k+1}-A_5$ and $A_k-A_4$ instead.
}
\label{bonds}
\begin{center}
\includegraphics[scale=.45]{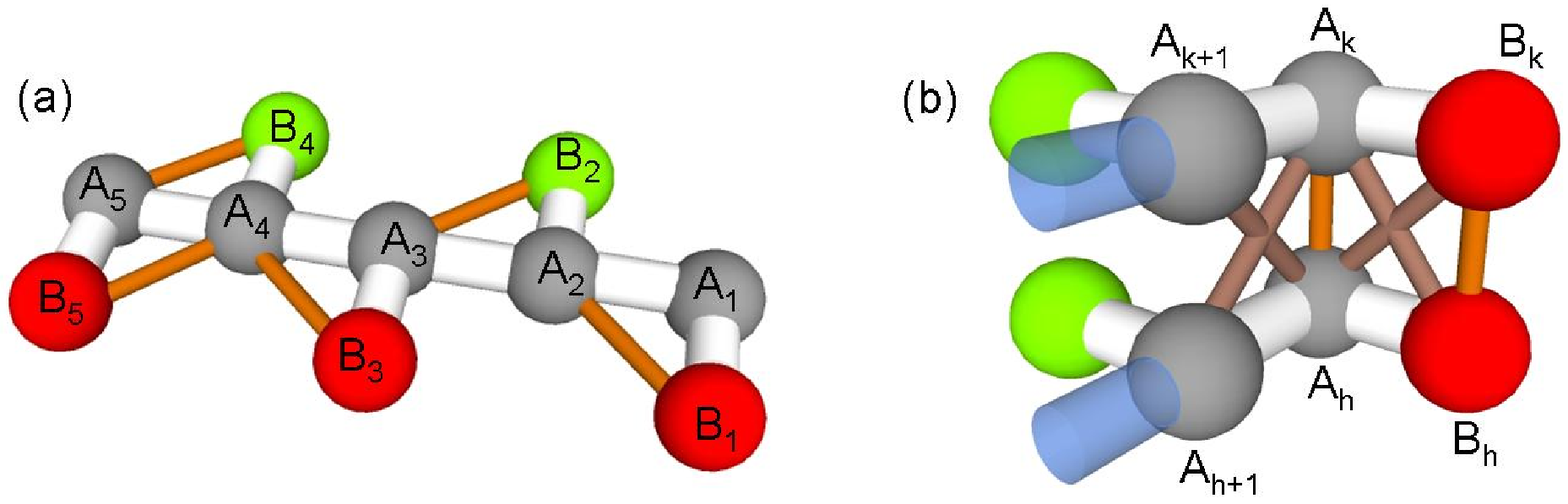}
\end{center}
\end{figure}

We will now describe the details of the model. For simplicity,
there are only two types of interactions:
\beqn
V(r,\kappa,\sigma,d) &=&  \left\{
\begin{array}{cl}
\frac{\kappa}{d^2}(r-\sigma)^2 -\kappa
& \sep |r-\sigma| < d
\\
0
& \sep {\rm otherwise}
\end{array}
\right.
\\
W(r,\kappa,\sigma,  d) &=&  \left\{
\begin{array}{cl}
\kappa\left[
\left(\frac{\sigma}{r}\right)^{12}-2
\left(\frac{\sigma}{r}\right)^{6}
\right]
& \sep r < d
\\
0
& \sep {\rm otherwise}
\end{array}
\right.
\ .
\eeqn
Namely, $V$ is a harmonic potential with a sharp cutoff at $d$ (in units of nm) and $W$ is the Lennard-Jones potential again with a cutoff at $d$, $\sigma$ (in units nm) denotes the minimum in the potential well and  $\kappa$
controls the depth of the potential well (in units of $k_BT$). Note the discontinuities in the slopes of the potentials at the cutoffs. We believe that these discontinuities are unimportant in our Langevin simulations due to the greater magnitude of perturbation from thermal fluctuation .

The whole system is modeled by pairwise interactions consisting of a linear sum of a set of potentials $V$ and $W$, which can be categorized into two classes: bonded interactions and non-bonded interactions.

\subsection{Bonded interactions}
Bonded interactions refer to the bonds within a peptide in order to provide it with the structural constraints that mimic a peptide (c.f. \fig \ref{bonds}). The parameters of the interactions are shown in \tab Ia.
For example, the total force acting on the bead $A_1$ due to bonded interactions is:
\beqn
\nonumber
&& \vnabla_{A_1} \Big[ V(r_{A_1A_2},40,1/2, \infty)+V(r_{A_1B_1},40,1/2, \infty)
\\
&+ &
 V(r_{A_1A_3},10,1, \infty)\Big] \ ,
\eeqn
where $\vnabla_\alpha \equiv \vnabla_{\br_{\alpha}}$ and $r_{\alpha \beta} = | \br_\alpha -\br_\beta|$ with $\br_\alpha$ referring to the position of the $\alpha$ bead. The first term fixes the inter-amino-acid distance to be about 0.5nm and the second term dictates that the distance between the peptide backbone and the side chain to be about 0.5nm. The second term above gives longitudinal rigidity to the peptide.

\begin{table}
\label{tables}
\caption{(a) The parameters employed in the bonded interactions. The potential is of the form $V$ with $d=\infty$ and the values in the entries correspond to $(\kappa, \sigma)$.
(b) The parameters for $U$-type non-bonded interactions. The potential is of the form $W$ and the values in the entries correspond to $(\kappa, \sigma, d)$.
(c) The parameters for $D$-type non-bonded interactions. The potential is of the form $V$ and the values in the entries correspond to $(\kappa, \sigma, d)$. Note that these interaction potentials are only switched on when all six distances are within the cutoffs.
}
\begin{center}
\includegraphics[scale=.4]{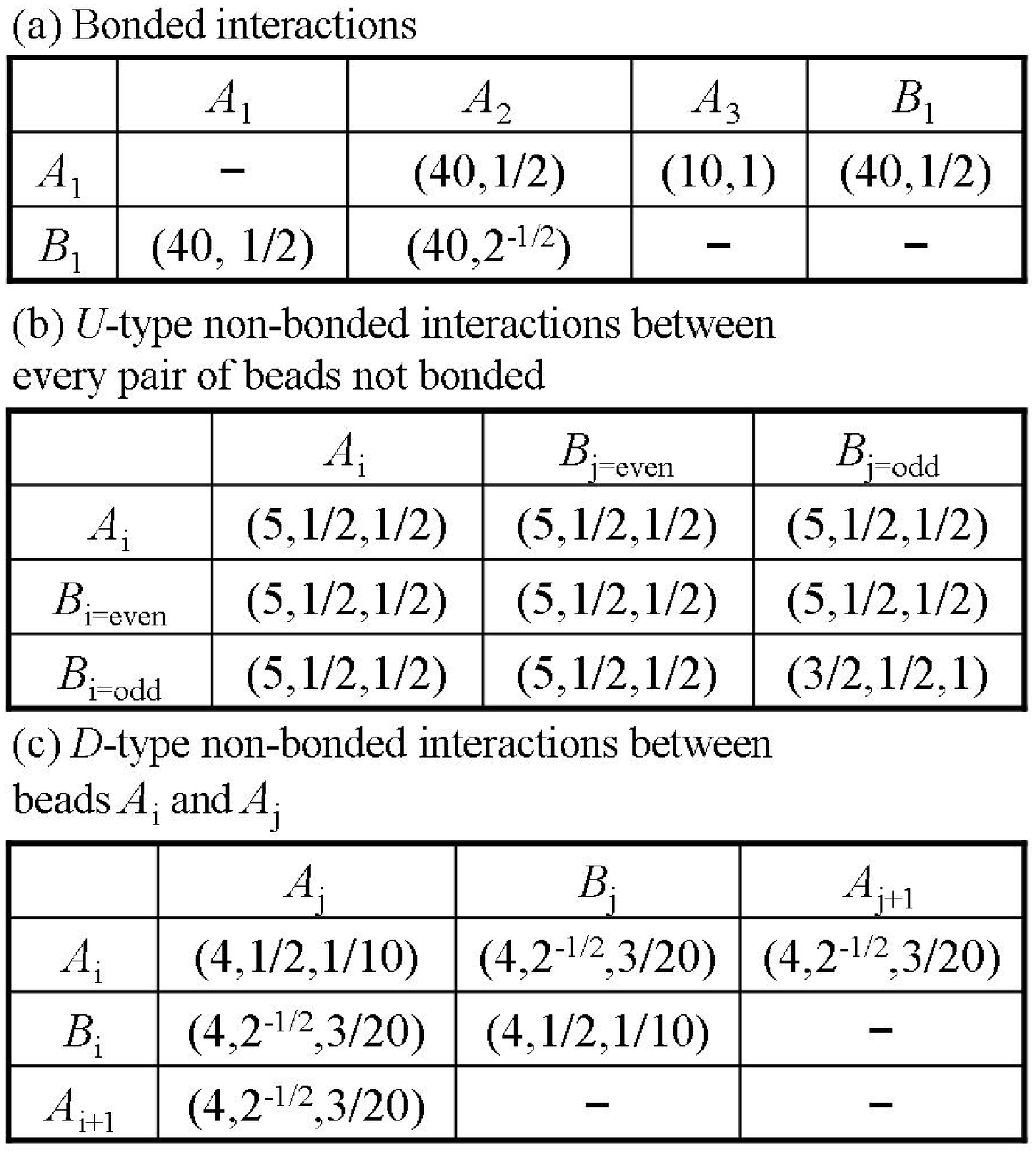}
\end{center}
\end{table}

\subsection{Non-bonded interactions}
Besides the bonded interactions, there are also non-bonded interactions between the beads. The first type is undirectional and we call them $U$-type interactions.

\subsubsection{$U$-type interactions}
These interactions include steric constraints or attractive interactions (only between pairs of hydrophobic beads, i.e., $B_{i = {\rm odd}}$) between every pair of beads, except for pairs already under bonded interactions. These effects are manifested by the the Lennard-Jones potential with different cutoffs, $d$: $d=\sigma$ for the case of pure repulsion, and $d> \sigma$ for the case of long range attraction with short range repulsion. The parameters employed in the simulations are shown in \tab Ib. For instance, the interaction potential between two red beads,  $B_i$ and $B_j$,  is $V(r_{B_iB_j},1,1/2,1)$.

\subsubsection{$D$-type interactions}
We use the term $D$-bonds to refer to the directional interactions between peptides. The $D$-bonds are meant to mimic the cross-beta sheet hydrogen bonds.
We will model the directional elements by a sum of six harmonic potentials $V$ (c.f. \fig \ref{bonds}b). This way of modeling directionality in hydrogen bonding is akin to the works employing discrete molecular dynamics simulations to study  amyloid formation in the literature \cite{Nguyen_PNAS04}. Note that the potentials are only switched on when all six pairwise distances concerned are within their respective cutoffs. In other words, the total interaction potential for a $D$-type bond between $A_i$ and $A_j$ is:
\beqn
\nonumber
&&
\theta \Big[ V(r_{A_iA_j},5,1/2,1/10)+
V(r_{A_iB_j},5,2^{-1/2},3/20)
\\
\nonumber
&+&
V(r_{A_iA_{j+1}},5,2^{-1/2},3/20)
+V(r_{B_iA_j},5,2^{-1/2},3/20)
\\
\nonumber
&+&
V(r_{B_iB_j},5,1/2,1/10)+
V(r_{A_{i+1}A_{j}},5,2^{-1/2},3/20) \Big] \ ,
\eeqn
where $\theta=1$ when all of the six distances are within their respective cutoffs, and $\theta=0$ otherwise.

\subsection{Simulation procedure}
We performed Langevin dynamics simulations on our system. Namely, the $\alpha$ bead follows the updating rule \cite{Thijssen_B07}:
\beq
\tri \br_\alpha = \frac{1}{\gamma} \vnabla_\alpha U_{\rm total} (\{\br\})\tri t+\sqrt{\frac{2k_BT}{\gamma}\tri t}\
\vec{\eta}_\alpha \ ,
\eeq
where $\vec{\eta}_\alpha$ represents Gaussian noise with zero mean and variance one in 3D,  $\gamma$ is the friction coefficient for each bead, and $U_{\rm total}$ is the sum of all pairwise interactions in the model. The relevant parameters are shown in \tab II.

\begin{table}
\label{table2}
\caption{
A comparison between the parameters employed in the simulations measure (middle column) and the corresponding values measured experimentally (right column). Note that the hydrophobic strength and the hydrogen bond strength in the middle column depicts the enthalpies defined in the simulations while the values in the right columns are measured in free energies.
We note that the qualitative nature of our conclusion stays the same when the hydrophobic strength (hydrogen bond strength) are varied around the presented values by twenty (ten) percents above and below.
}
\begin{center}
\begin{tabular}{|l|c|c|} \hline
Properties &
Simul. & Exp'l
\\ \hline
Friction coefficient per bead, $\gamma$ ($k_BT$ps/nm$^2$) &  1000 & $\sim 1000$ \cite{Skjaeveland_EPJE00}
\\ \hline
Inter-amino-acid distance (nm) & 0.5 & 0.35 \cite{Daune_B99}
\\ \hline
Hydrophobic interaction strength ($k_BT$) & 1.5 & 1-4 \cite{Jackson_B06}
\\ \hline
Hydrogen bond strength ($k_BT$) & 4 & $\sim 2.3$ \cite{Jackson_B06}
\\ \hline
Time increment,  $\tri t$ (fs) & 5.6 &  --
\\ \hline
\end{tabular}
\end{center}
\end{table}

\begin{figure}
\caption{
A snapshot of one simulation run in progress. The fibril is put in the middle of the simulation box and the monomer is diffusing towards it.
The fibrillar axis is along the $z$-axis, the cross-beta sheets are along the $x$-axis, and the
coordinates of the $A_3$ bead for the peptides are $(-0.25,0.6,k/2-1)$ and $(0.25,0.6,k/2-1.25)$, $k=0, \ldots 4$. The corners of the simulation box are at $x,y,z = \pm 3$.
}
\label{Nereus}
\begin{center}
\includegraphics[scale=.4]{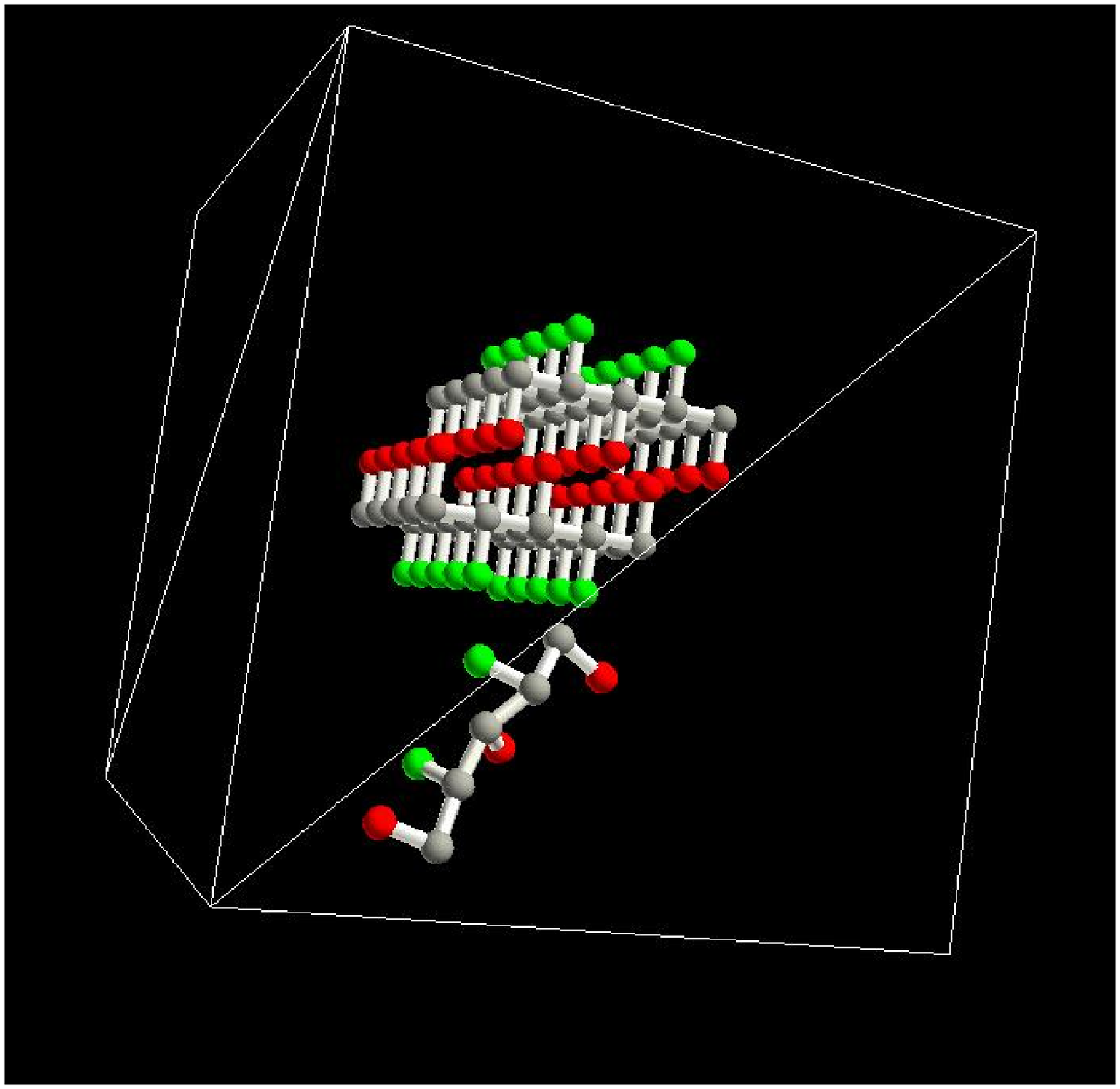}
\end{center}
\end{figure}

Simulations are done with one fibril segment and one monomer in a cubic box 6nm on a side (therefore, the monomer concentration $[m]$ and fibril concentration $[f]$ is 7.7mM). A fibril segment consisting of ten 5-amino-acid peptides are placed at the center of the box. The fibril is constructed by hand and consists of two-layer of cross-beta sheet structure as depicted in \fig \ref{Nereus}. The fibril is held fixed, i.e., the peptides within it are completely frozen throughout the simulation. At time zero, a monomeric peptide is placed at the corner of the box and the simulation is stopped when the free monomer has all of the five $D$-bonds formed with the fibril. Note that there are four possible locations for the added monomer to bind to as the fibril has two ends and there are two cross-beta sheets.

Throughout the run, we record the time when a change in the number of $D$-bonds  between the monomeric peptide and the fibril happens. This allows us to construct a time series describing the temporal evolution of the elongation process. We will now describe how the time series is analyzed.

\section{Data analysis}
\label{data}
\begin{figure}
\caption{
A segment of the data from the simulations. (a) The original time series consists of the times when the number of $D$-bond is changed. This is then transformed into a time series on the transitions of the set of states $\{S\}$ (b). The procedure of transformation is described in the text.
}
\label{series}
\begin{center}
\includegraphics[scale=.4]{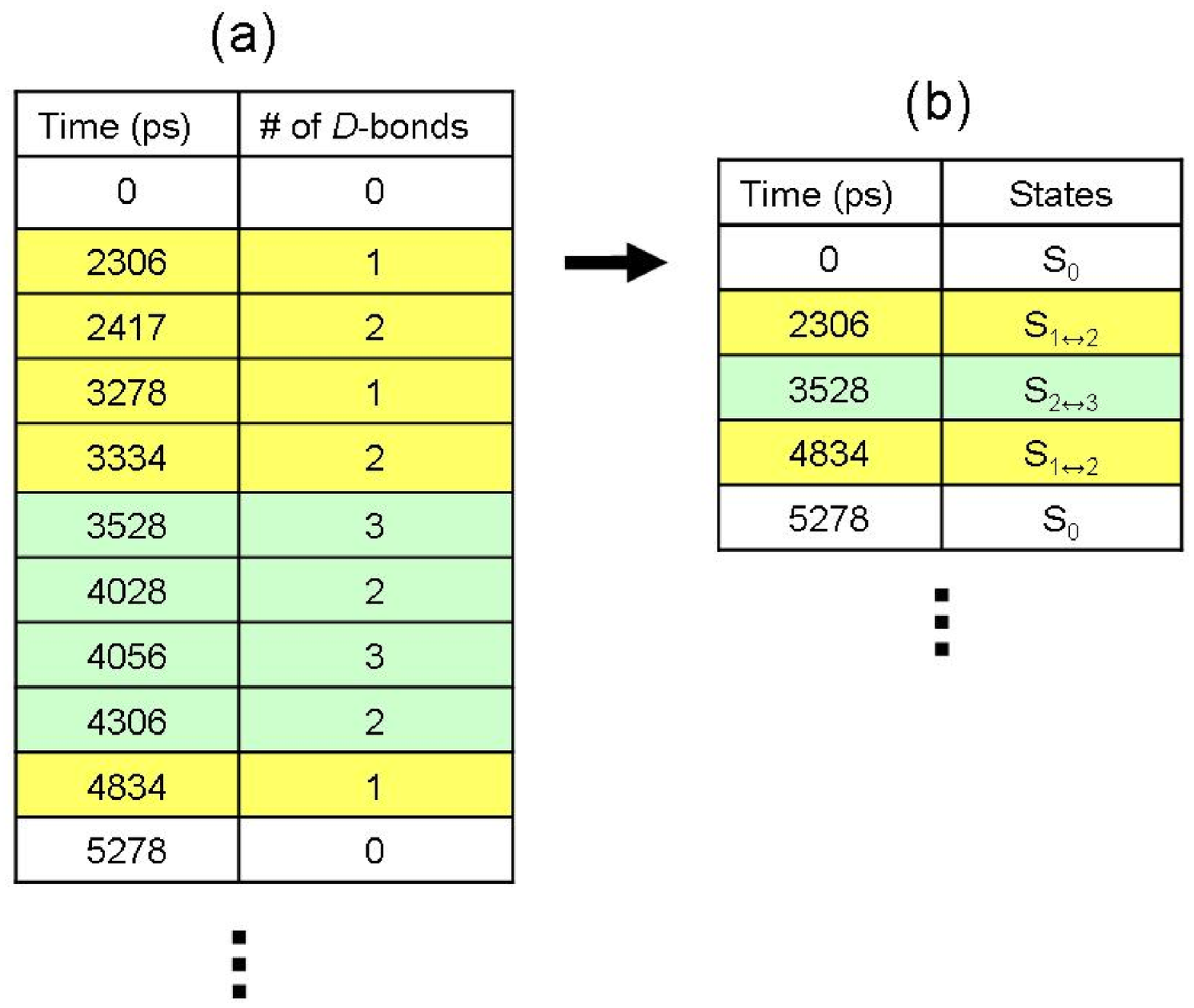}
\end{center}
\end{figure}

\begin{figure}
\label{stat}
\caption{
The numbers of transitions within the set of states $\{S\}$. Note the $ij$-entry denotes the number of transitions from state $S_{i-1}$ to $S_{i-1 \leftrightarrow j}$.
}
\begin{center}
\includegraphics[scale=.3]{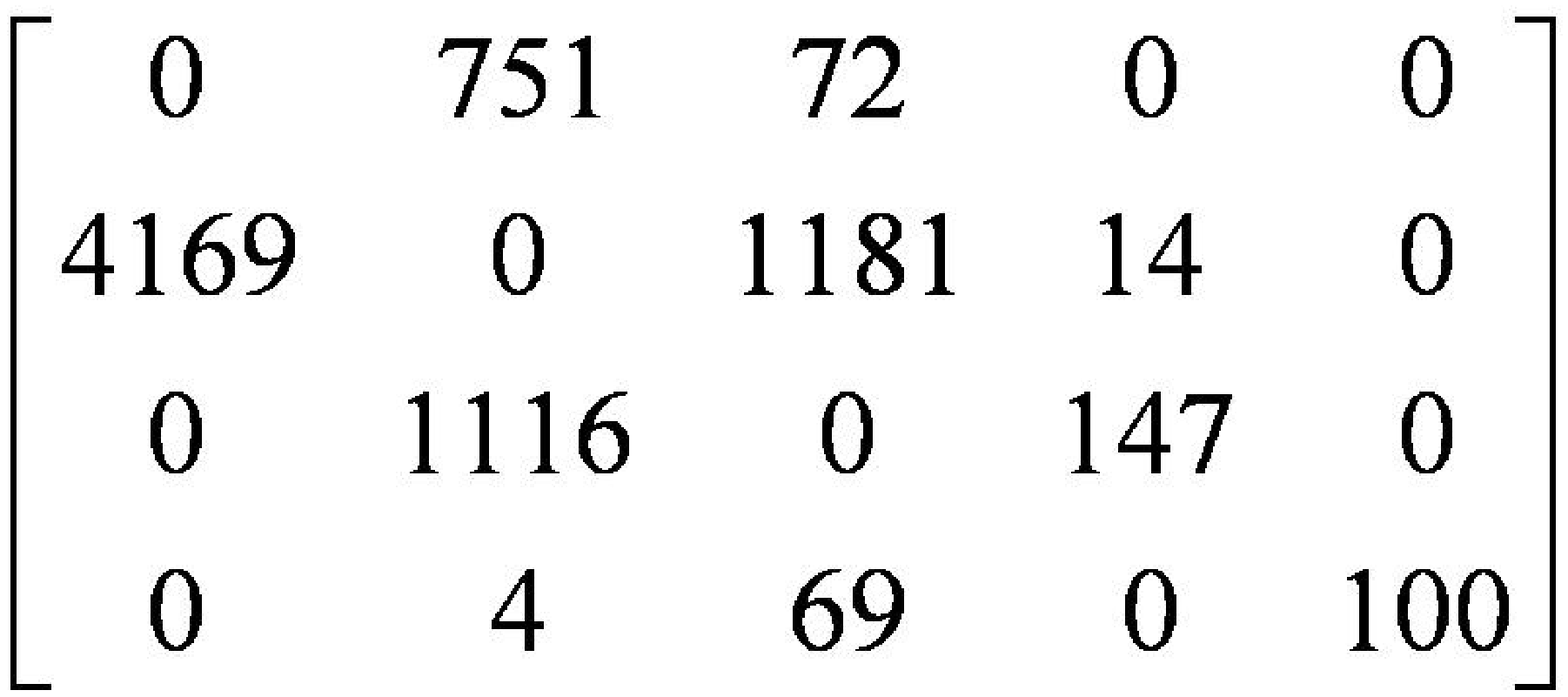}
\end{center}
\end{figure}

\begin{figure}
\caption{
The transition rates for the set of states $\{S\}$ constructed from the time series as described in the text. The units are in ps$^{-1}$.}
\label{rate_pic}
\begin{center}
\includegraphics[scale=.4]{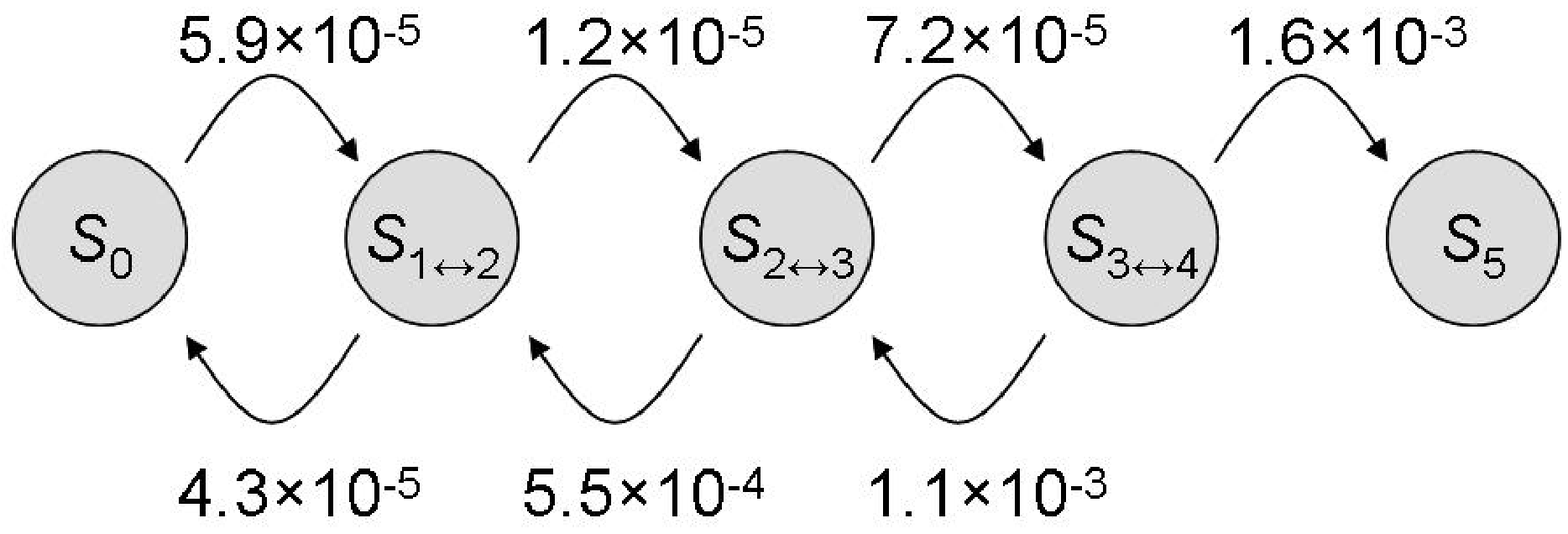}
\end{center}
\end{figure}

To comprehend our simulation results, we adopt a coarse-grained picture of the dynamics of fibril elongation. Specifically,  we partition the phase space of a monomer in the process of fibrillization into a number of discrete states, and aim to approximate the dynamical picture by a series of jumps between neighboring states by Markovian processes. The desire to have a Markovian representation is an attempt to view the dynamics through a familiar mechanism.
To find a sensible definition for the set of discrete states, we firstly simplify the dynamical picture by recoding the time whenever a $D$-bond is formed or destroyed. This gives an array consisting of the times and the numbers of $D$-bonds between the monomer and the fibril as shown in \fig \ref{series}a. By inspection of the dataset, it is apparent that the time series consists of segments of long periods within which there are a lot of rapid back and forth transitions between having $k$ and $k+1$ $D$-bonds. This is a clear sign of temporal correlation and since our desire is to approximate the process with a memoryless kinetic mechanism, we will  partition the configuration space of our system into five discrete states designated by: $S_0$, $S_{1\leftrightarrow 2}$, $S_{2\leftrightarrow 3}$, $S_{3\leftrightarrow 4}$ and $S_{5}$, where $S_0$ refers to having no $A$-bonds between the monomer and the fibril, $S_{k\leftrightarrow k+1}$ refers to the state where the number of $D$-bonds flickers between $k$ and $k+1$, and $S_5$ refers to the fully aligned state for the monomer.
With these newly defined states, a new time series recording the transitions between them can be constructed (c.f. \fig \ref{series} and \fig \ref{stat}).
We now assume that all the transition events are drawn from Independent and Exponential Distributions. Given the property that the minimum of two exponential random variables is again exponentially distributed with rate equal to the sum of the two original rates, we are able to decouple the individual rate for each transition event from the time series (c.f. footnote 42). The results are shown in \fig \ref{rate_pic}.

\begin{figure}
\caption{
(a) The dock-and-convert free energy landscape picture of the elongation process proposed in \cite{Kusumoto_PNAS98,Esler_Biochem00,Massi_Proteins00,Scheibel_PNAS04,Nguyen_PNAS07}.  $S_0$ denotes the initial state, i.e., a free monomer, and $S_5$ denotes the final state with monomer being part of the fibril.
(b) The free energy landscape picture advocated in this work. The schematic underneath the landscape picture depicts one particular conformation in the $S_{2\leftrightarrow 3}$ state trapped due to misalignment. Note that there are three $D$-bonds formed with the fibril and the cloud encloses the dangling end with two unbound $A$ beads.
}
\label{landscape}
\begin{center}
\includegraphics[scale=.4]{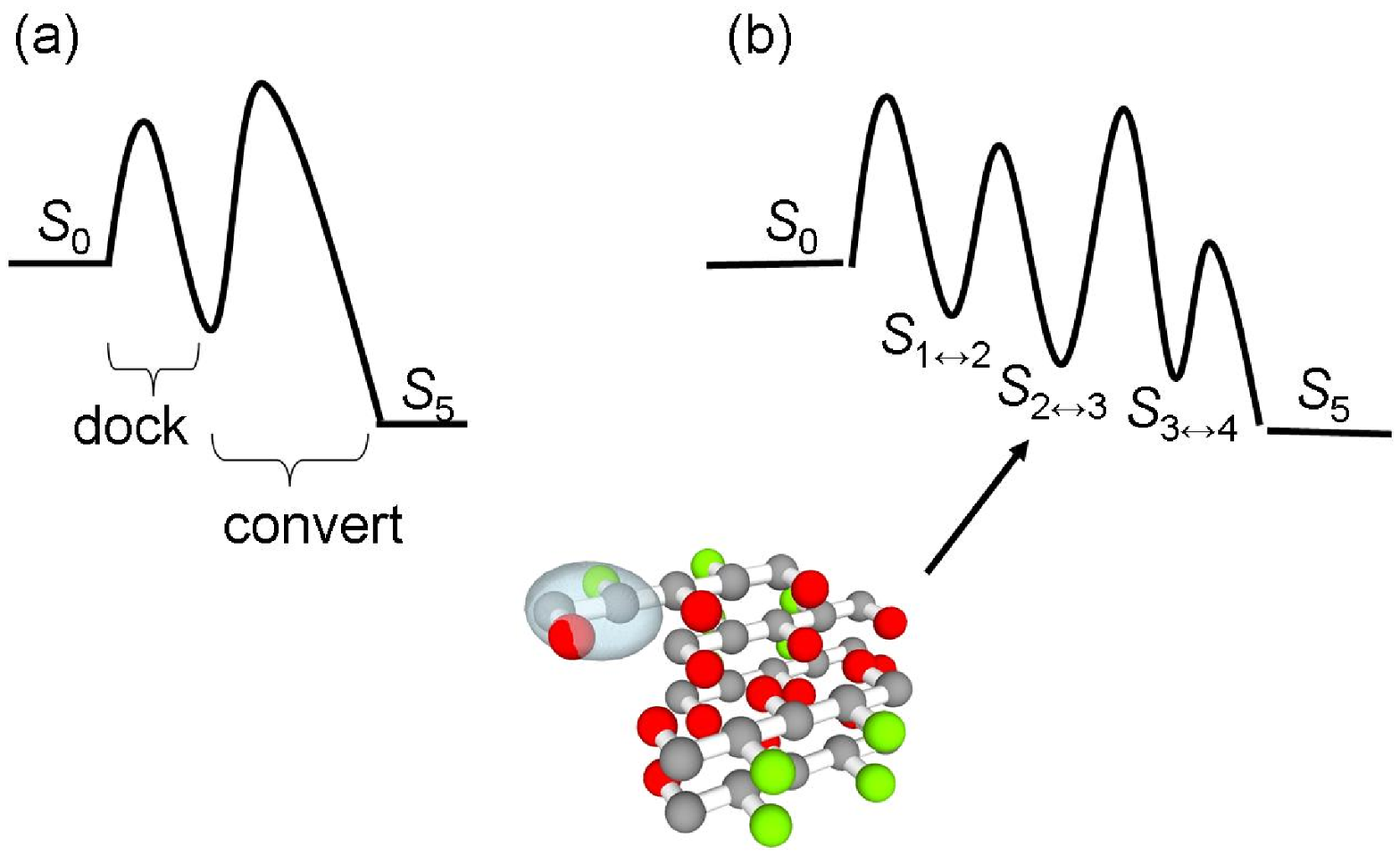}
\end{center}
\end{figure}

\section{Discussion}
\label{discussion}
The diffusion constant of a monomer is measured to be $1.1 \times 10^{-4}$nm$^2$/ps (plots not shown). Since the combined binding area for the monomer and fibril's ends is about 5nm$^2$, we expect that the collision frequency in our system (with $[m]=[f]=7.7$mM) to be about $5\times 10^{-5}$. We  therefore conclude that the initial binding event (from state $S_0$ to state $S_{1\leftrightarrow 2}$) is well described by a diffusion controlled reaction.

Besides the transition rate between $S_0$ and $S_{1\leftrightarrow 2}$, we can see that many of the forward transition rates are of the same order of magnitude as the first binding rate. This demonstrates that the elongation process is not first-order, but rather dominated by frustrations for the monomer to find the correct configuration to become fully part of the fibril, i.e., state $S_5$ (c.f. \fig \ref{landscape}).  We note that this qualitative picture stays the same when the hydrophobic strength (hydrogen bond strength) are varied around the presented values by twenty (ten) percents above and below.
This is the main result of this work. 

To have a conceptual feeling for how a rugged energy landscape picture would affect the elongation process, we first look at the work by Zwanzig \cite{Zwanzig_PNAS88}, which demonstrates that: if a particle is diffusing over a 1D rugged landscape such that the fluctuation in potential energy is Gaussian distributed with zero mean and standard deviation $\epsilon$, then the motion of the particle can be effectively described by ordinary diffusion with a re-defined diffusion constant, $D^*$, of the form:
\beq
\label{D_def}
D^*=D \exp[-(\epsilon/k_BT)^2]
\eeq
where $D$ is the original diffusion coefficient.

\begin{figure}
\caption{
A schematic diagram depicting the scenario where the elongation process is viewed as a diffusion process over a rugged energy landscape in 1D. The linear dimension denotes the `reaction coordinate' and the position $x_0$ indicates the location when the monomer is first bound to the fibril's end.
}
\label{new_model}
\begin{center}
\includegraphics[scale=.45]{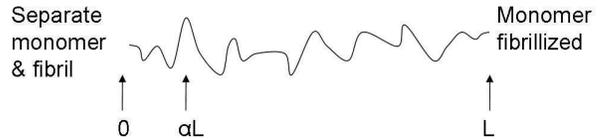}
\end{center}
\end{figure}

Let us consider the elongation process as a drift-diffusion process on a rugged energy landscape (c.f. \fig \ref{new_model}). Adopting the idea of Zwanzig mentioned above \cite{Zwanzig_PNAS88}, we account the ruggedness by redefining the diffusion constant as in \eq \ref{D_def}. In other words, the probability distribution, $p(x,t)$, of the state of the system (represented by the reaction coordinate $x$) follows the differential equation below:
\beq
\pp_t p(x,t) = D\pp_x^2 p(x,t) - v\pp_x p(x,t) \ .
\eeq
where $D$ is the renormalized diffusion constant that takes the ruggedness into account, and $v$ is the drift produced by the free energy descent that drives the monomer to become fibrillized. The differential equation is supplemented by the boundary condition $p(0,t)=p(L,t)=0$ where the left boundary depicts monomer detachment from the fibril's end and the right boundary depicts completion of the fibrillization process (c.f. \fig \ref{new_model}).
We will now assume that the initial condition is a delta function located at $\alpha L$, i.e.,  $p(x,t=0) = \delta(x- \alpha L)$, such that $\alpha$ is small. If $v$ is negligible, i.e., when the free energy drive for fibrillization is negligible, the ratio of monomers exiting at the left boundary (becoming detached) and exiting at the right boundary (becoming fibrillized) is
proportional to $\alpha$ \cite{Farkas_JPA01}. Let us now use the number of hydrogen bonds again as a very crude estimate for the reaction coordinate. For the case Abeta peptides, the total number of hydrogen bonds is likely to be in the order of 20 \cite{Petkova_PNAS02} so if we take the initial location as having one hydrogen bond formed with the fibril, $\alpha \sim 1/20$. In other words, according to this diffusion-on-rugged-landscape model, only about 20 monomers would be recycled before one of them is fibrillized, as compared to the $10^8$ monomers predicted by the two-step model depicted in \eq \ref{scheme}.

Our models also provides the following insights on the elongation process:
\begin{enumerate}
\item
During the period of conversion, the monomer will go through a lot of  different conformations, and there is not a specific conformation that acts as the typical conformation before fibrillization.
\item
Since the conversion step is slow, the interactions between multiple monomers at the fibrils' ends should be important, and this propensity for monomers' interactions may also serve to promote oligomers formation. 
One would also expect that multi-monomer interactions would induce more ruggedness into the landscape picture.
\item
Since elongation rates are determined by the form of the energy traps, it is intuitive to expect that the more uniform the amyloid-forming peptide's sequence is, the slower the elongation rate. This is because primary sequence with many identical side chains would promote misalignment binding and as a result, enhance the ruggedness of the energy landscape. In other words, the complexity of the primary sequence may serve as a factor in elongation rate prediction (c.f. \cite{Yoon_ProtSci04,
Fernandez_NatBiotech04,Tartaglia_ProtSci05,Galzitskaya_PLoSCompBiol06,DuBay_JMB04,Pawar_JMB05}).
\end{enumerate}

\section{Conclusion}
We have studied the elongation process of amyloid fibril by performing Langevin simulations on a toy model of peptides. By projecting the elongation process onto a set of discrete states, a rugged energy landscape picture emerged, which indicates that monomer-fibril interaction is prolonged in the course of elongation. Our findings also
suggest that the complexity of an amyloid forming peptide, as measured for instance by how diverse the amino-acid compositions are, may serve as a predictor of the fibril elongation rate. These conclusions can be tested with current experimental techniques.

\begin{acknowledgements}
The simulations are performed with Nereus System \cite{nereus}, a distributive grid computing system developed in Particle Physics (Oxford).
CFL thanks the Glasstone Trust (Oxford) and Jesus College (Oxford), JL the John Fell OUP Fund, for financial support. LJ and DJV acknowledge the generous support of Synatica Ltd, a spin-out company of the University of Oxford.
\end{acknowledgements}

\bibliography{chiufanlee,books}
\end{document}